\documentclass[useAMS,usenatbib,usegraphicx,letters]{mn2e}
\usepackage{graphicx, txfonts, color, soul,ulem, multirow}

\title[The shocking transit of WASP-12b]{The shocking transit of WASP-12b: Modelling the observed early ingress in the near ultraviolet}
\author[Llama et al.]{J. Llama$^1$\thanks{E-mail: jl386@st-andrews.ac.uk}, K. Wood$^1$, M.~Jardine$^1$, A.~A.~Vidotto$^1$,  Ch.~Helling$^1$, L. Fossati$^2$,\newauthor  C.~A.~ Haswell$^2$\\
$^1$SUPA, School of Physics \& Astronomy, University of St Andrews, North Haugh, St Andrews,  KY16 9SS, UK\\
$^2$ Department of Physics and Astronomy, Open University, Walton Hall, Milton Keynes MK7 6AA
}

\begin{document}

\date{Accepted ??. Received ??; in original form ??}

\pagerange{\pageref{firstpage}--\pageref{lastpage}} \pubyear{2011}

\maketitle

\label{firstpage}

\begin{abstract}
Near ultraviolet observations of WASP-12b have revealed an early ingress compared to the optical transit lightcurve. This has been interpreted as due to the presence of a magnetospheric bow shock which forms when the relative velocity of the planetary and stellar material is supersonic. We aim to reproduce this observed early ingress by modelling the stellar wind (or coronal plasma) in order to derive the speed and density of the material at the planetary orbital radius. From this we determine the orientation of the shock and the density of compressed plasma behind it. With this model for the density structure surrounding the planet we perform Monte Carlo radiation transfer simulations of the near UV transits of WASP-12b with and without a bow shock. We find that we can reproduce the transit lightcurves with a wide range of plasma temperatures, shock geometries and optical depths. Our results support the hypothesis that a bow shock could explain the observed early ingress.  
\end{abstract}

\begin{keywords}
planet-star interactions --- planets and satellites: individual (WASP-12b) --- planets and satellites: magnetic fields --- stars: coronae --- stars: winds, outflows
\end{keywords}

\section{Introduction}
 The detection of hot-Jupiters, which are found to be orbiting extremely close to their host star with periods of a few days, is continuing to challenge our theories of planet formation and evolution \citep{Watson:2011p826}.  Transit surveys such as SuperWASP  provide great insight into the properties of extrasolar planets and their host star \citep{Seager:2003p810}. As the number of detected planets has increased, work has started on characterising their atmospheres \citep{Madhusudhan:2011p830,Croll:2011p831,Helling:2011p833} including  cloud coverage and temperature structure \citep{VidalMadjar:2011p825}.  
 
Following on from the identification the hot-Jupiter WASP-12b (\citealt{Hebb:2009p806}), recent {\it HST} observations indicate an early ingress of its near ultraviolet transit compared to the optical data  \citep{Fossati:2010p838}. There have been several attempts to explain such an early ingress by assuming the presence of absorbing material around the planet \citep{Lai:2010p808, Vidotto:2010p809}.  One possible explanation for the presence of an asymmetry is that heavily irradiated gas giants such as WASP-12b can fill and even overflow the planetary Roche lobe \citep{Gu:2003p816, Ibgui:2010p817, Li:2010p818} resulting in mass transfer from the planet to the star \citep{Lai:2010p808}.

 One further explanation for this observed asymmetry is the presence of a bow shock. \cite{Vidotto:2010p809}  have shown that if the relative motion between the planetary and stellar coronal material is supersonic, then a bow shock surrounding the planet's magnetosphere could form. Such a shock could compress the local plasma to the densities required to reproduce the early ingress observed in the near UV lightcurves. 
 
 The possibility of detecting and characterising planetary bow shocks provides a new method to study exoplanetary magnetic fields. If the magnetic field of the star can be determined through, for example,  Zeeman-Doppler imaging techniques \citep{Donati:2009p824}, then the early ingress of the near ultraviolet transit can be used to place limits on the exoplanetary magnetic field \citep{Vidotto:2011p803}. The magnetic field is believed to provide a key role in shielding the planetary atmosphere from energetic particles.  

In this Letter, we present theoretical light curves for a planet surrounded by a magnetospheric bow shock transiting its host star. Our aim is to test the hypothesis that the early ingress of the near UV transit lightcurve can be explained by the presence of such a bow shock.
 
\section{The Model}
\begin{figure}    \centering
   \includegraphics[height=2.8in]{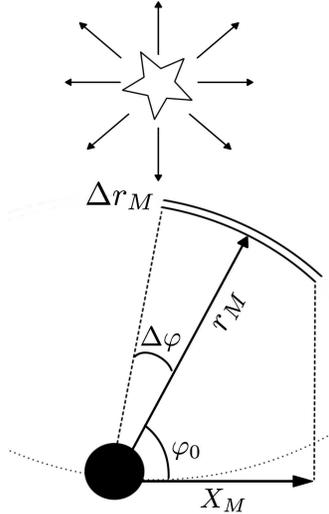} 
   \caption{Schematic of the shock geometry as viewed along the stellar rotation axis (not drawn to scale). The shock normal makes an angle $\varphi_0$ to the direction of motion of the planet. The distance from the planet to the shock, $r_M$, is determined primarily by the strength of the planetary magnetic field. $X_M$ denotes the maximum distance between the planet and the projected lateral extent of the shock. This distance depends on the chosen values for $r_M, \varphi_0$ and $\Delta\varphi$.   $\Delta r_M$ is the region of compressed stellar material behind the shock.}
   \label{fig:schematic}
\end{figure}

If the relative velocity of the planetary and the stellar coronal material is supersonic then a bow shock could form (see Figure \ref{fig:schematic}). The interaction between planetary material and stellar material compresses the local plasma to produce a region of higher density plasma behind the shock. If the optical depth of the shocked material is high enough then starlight will be absorbed and produce an early ingress in the transit lightcurve. 

The angle between the shock normal and the orbital direction of the planet is given by $\varphi_0$. The value of $\varphi_0$ is determined by the relative velocity of the planetary and stellar coronal material. Figure 1 of \cite{Vidotto:2010p809} illustrates the scenarios leading to the various shock orientations. There are two limiting cases: an ``ahead-shock" ($\varphi_0\rightarrow0$) forms when the planet is embedded in the stellar corona and a ``dayside-shock" ($\varphi_0\rightarrow90^\circ$) forms when the radial wind velocity is very much greater than the relative azimuthal velocity of the planet. 
 
Here we use models for the stellar corona and wind \citep{Vidotto:2010p809,Vidotto:2011p803} to obtain a value for the plasma density at the planet. These models assume a typical solar base density of $n_0\sim10^8\textrm{cm}^{-3}$ \citep{Withbroe:1988p829} and either an isothermal hydrostatic corona or an isothermal thermally driven wind. We assume an adiabatic shock with a maximum compression ratio of 4.  For an isothermal corona of temperature $T$, the density of stellar material, $n_{\rm obs}$, at a planet of orbital radius $R_{\rm orb}$ is given by Equation 8 of \cite{Vidotto:2010p809}, 
\begin{eqnarray} 
	 \frac{n_{\rm obs}}{n_0}=\exp\left[\frac{GM_\star/R_\star}{k_BT/m}\left(\frac{R_\star}{R_{\rm orb}}-1\right) +\frac{2\pi^2R_\star^2/P_\star^2}{k_BT/m}\left(\frac{R_{\rm orb}^2}{R_\star^2}-1\right)\right].
\label{eqn:density}
\end{eqnarray}
For the stellar wind case we use mass conservation $(nu_rr^2=\textrm{const})$ and the momentum equation,
\begin{equation}
	\rho u_r\frac{\partial u_r}{\partial r}=-\frac{\partial p}{\partial r}-\rho\frac{GM_\star}{r^2}
\label{eqn:windode}
\end{equation}
to obtain values for the radial velocity, $u_r$ and density, $n_{\rm obs}$. The plasma density can then be converted into a density of fully ionized magnesium using the relation,
\begin{equation}
	n_{\rm MgII} = 4\times n_{\rm obs}\times \frac{n_{\rm Mg}}{n_{\rm H}},
\label{eqn:mgconversion}
\end{equation}
where, $n_{\rm Mg}/n_{\rm H}$ is the ratio of Magnesium number density to Hydrogen number density which is derived from the metallicity of the host star \citep{Hebb:2009p806}. For WASP-12, $n_{\rm Mg}/n_{\rm H}=6.76\times10^{-5}$  \citep{Vidotto:2010p809}. From this density we can then find bow shock geometries and orientations that fit the \textit{HST} observations of \cite{Fossati:2010p838}.

To investigate whether the model presented by \cite{Vidotto:2010p809} is able to reproduce the data from the near-UV observations, we use Monte Carlo radiative transfer calculations to produce simulated light curves. The parameters we adopt to match the WASP-12 system are: $M_p = 1.41 M_J$,  $R_p = 1.79 R_J$ (where $M_J$ and $R_J$ are the mass and radius of Jupiter), $M_\star = 1.35 M_\odot$ and $R_\star = 1.57 R_\odot$. The host star is a late F type and the planet orbits in the equatorial plane with an impact parameter $b= 0.36\,R_\star$  \citep{Hebb:2009p806}.  
 
 The shocked material is considered to be at a distance $r_M$ from the planet, with a thickness $\Delta r_M$ and an angular extent $2\Delta \varphi$. The projected lateral extent of the shock is dependent on $r_M,\varphi_0$ and $\Delta\varphi$. The maximum distance between the planet and the projected lateral extent of the shock, $X_M$, can take the following forms 
\begin{equation}
	X_M =
\left\{
	\begin{array}{ll}
		r_M   & \mbox{if } \varphi_0\leq\Delta\varphi \\
		r_M\cos(\varphi_0 -\Delta\varphi)& \mbox{if } \varphi_0>\Delta\varphi .
	\end{array}
\right.
\label{eqn:xm}
\end{equation}
\begin{table}
\centering
	\caption{Parameters used in our simulations. In all cases we use $X_M=5.5R_p$.  The columns are respectively: Model number;  Temperature of stellar plasma;  Angle of the shock normal;  Calculated value for the number density of MgII;   Required thickness of shocked material from our simulations to reproduce the early ingress;  Angular extent of the shock;  Calculated maximum optical depth for each model.}
 \begin{tabular}{@{}cccccccc@{}}
	\hline
	Model  & T & $\varphi_0$ & $n_\mathrm{MgII}$ &  $\Delta r_M $ & $\Delta\varphi$ & $\tau_\mathrm{max}$ \\
	& (MK) & $(^\circ)$ & $(\textrm{cm}^{-3})^*$   &$(R_p)$& $(^\circ)$ & $(R_p)$\\
	\hline
	\hline
	1A & 2 & 0 & 312 &  0.08 & 80 & 0.49\\
	1B& 2 & 34 & 192 &  0.32 & 40 & 0.44\\
	\hline
	2A & 2.5 & 0 & 745  &0.05 & 80 & 0.92\\
	2B & 2.5 & 42 & 385 &0.15 &40 &0.37\\
	\hline
	\multicolumn{8}{l}{$^*$ assuming a solar corona base density $n_0\sim10^8\textrm{cm}^{-3}$}
	\end{tabular} 
	\label{table:results}
\end{table}
\begin{figure*}    \centering
   \includegraphics[height=3in]{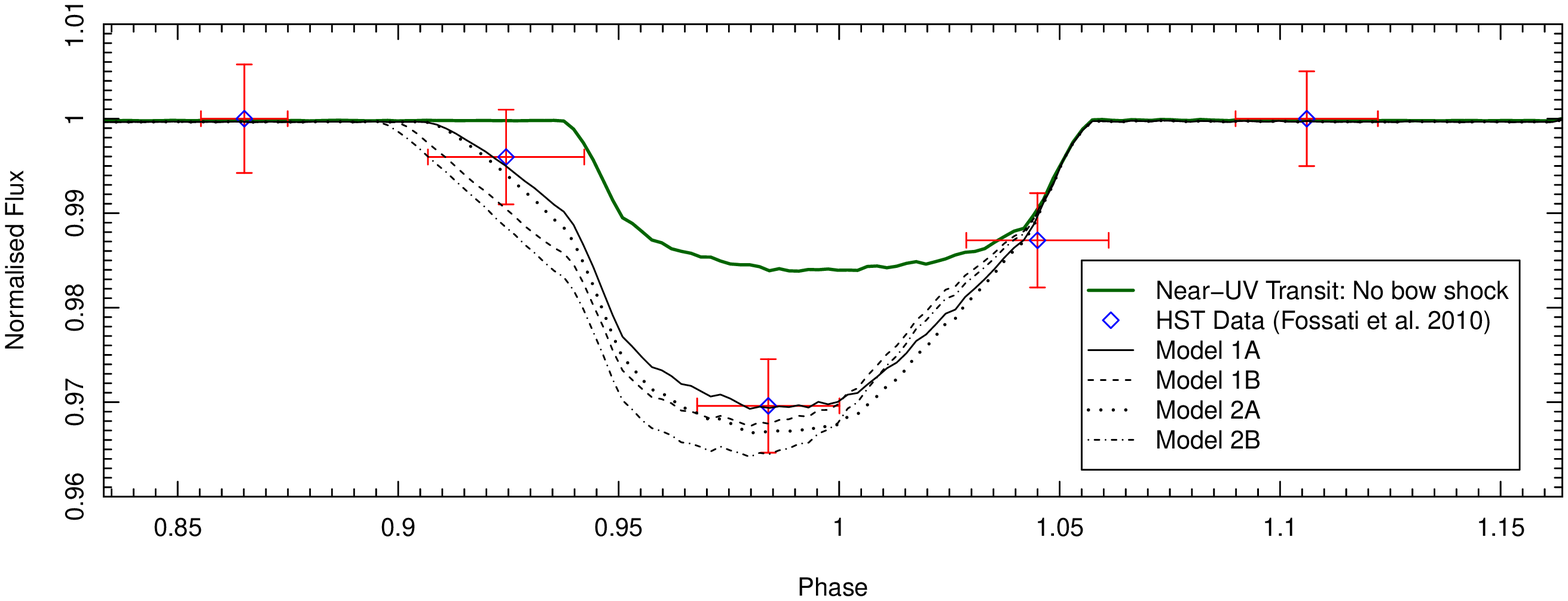} \\
   \vspace{-0.15in}
    \includegraphics[width=1.5in]{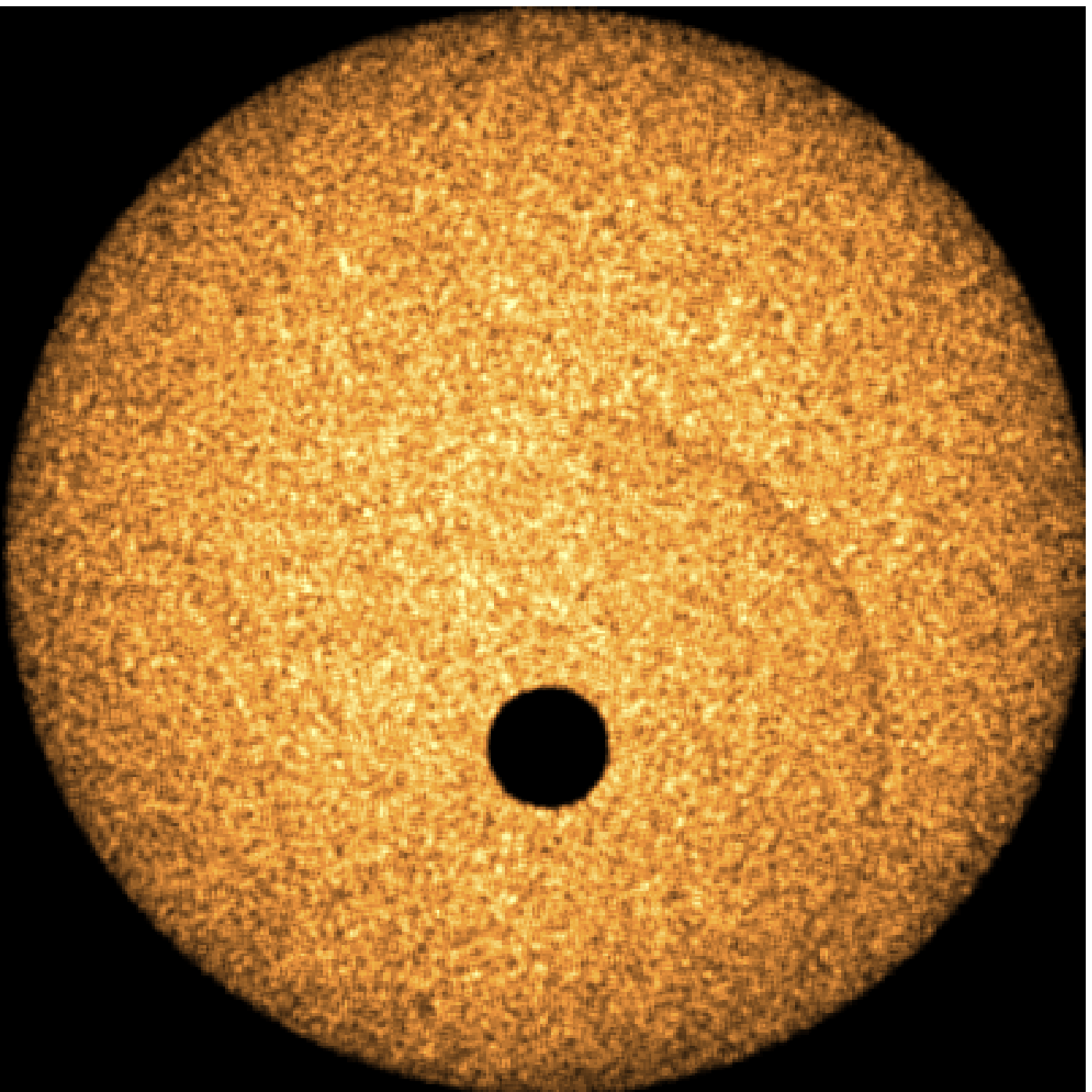}
    \includegraphics[width=1.5in]{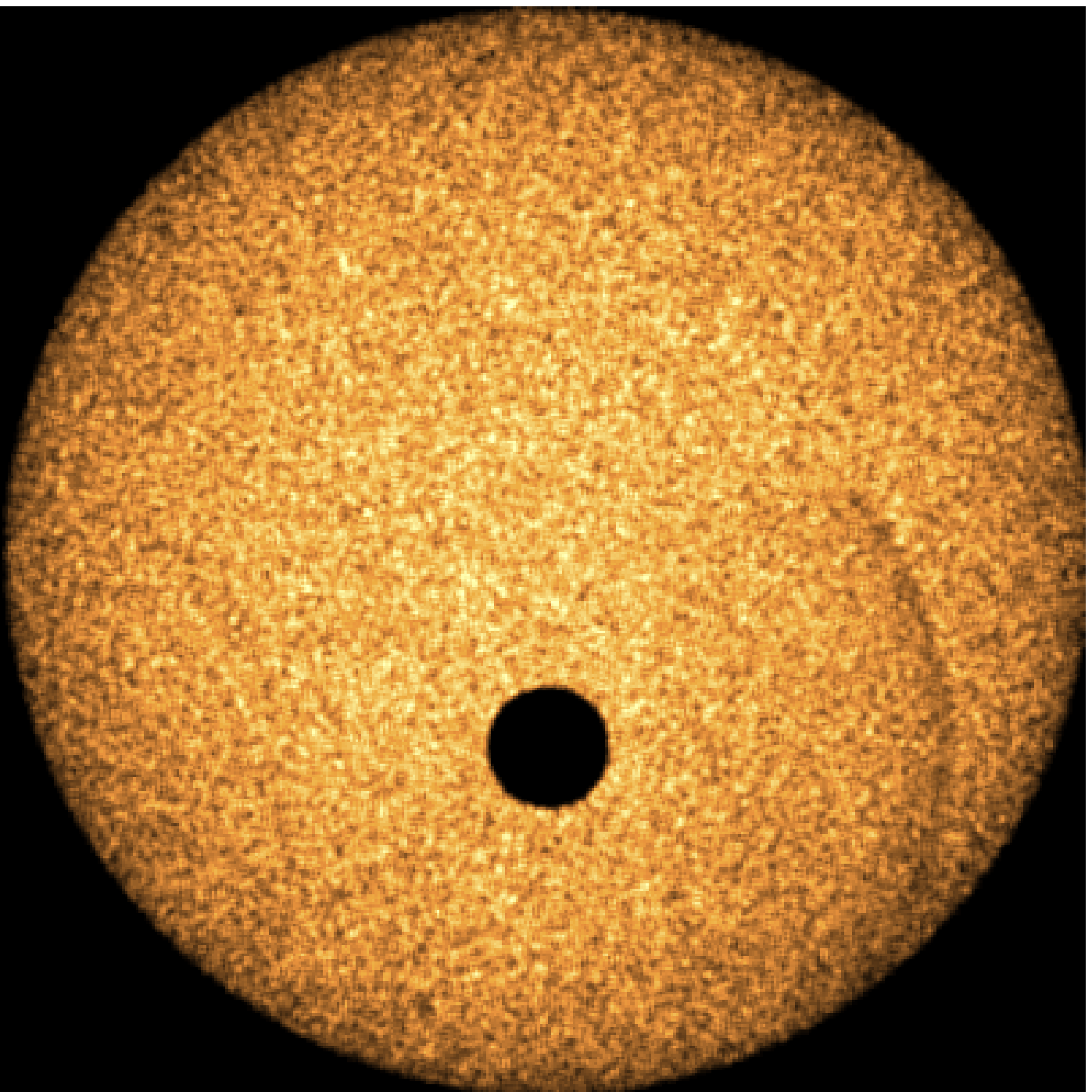}  
     \includegraphics[width=1.5in]{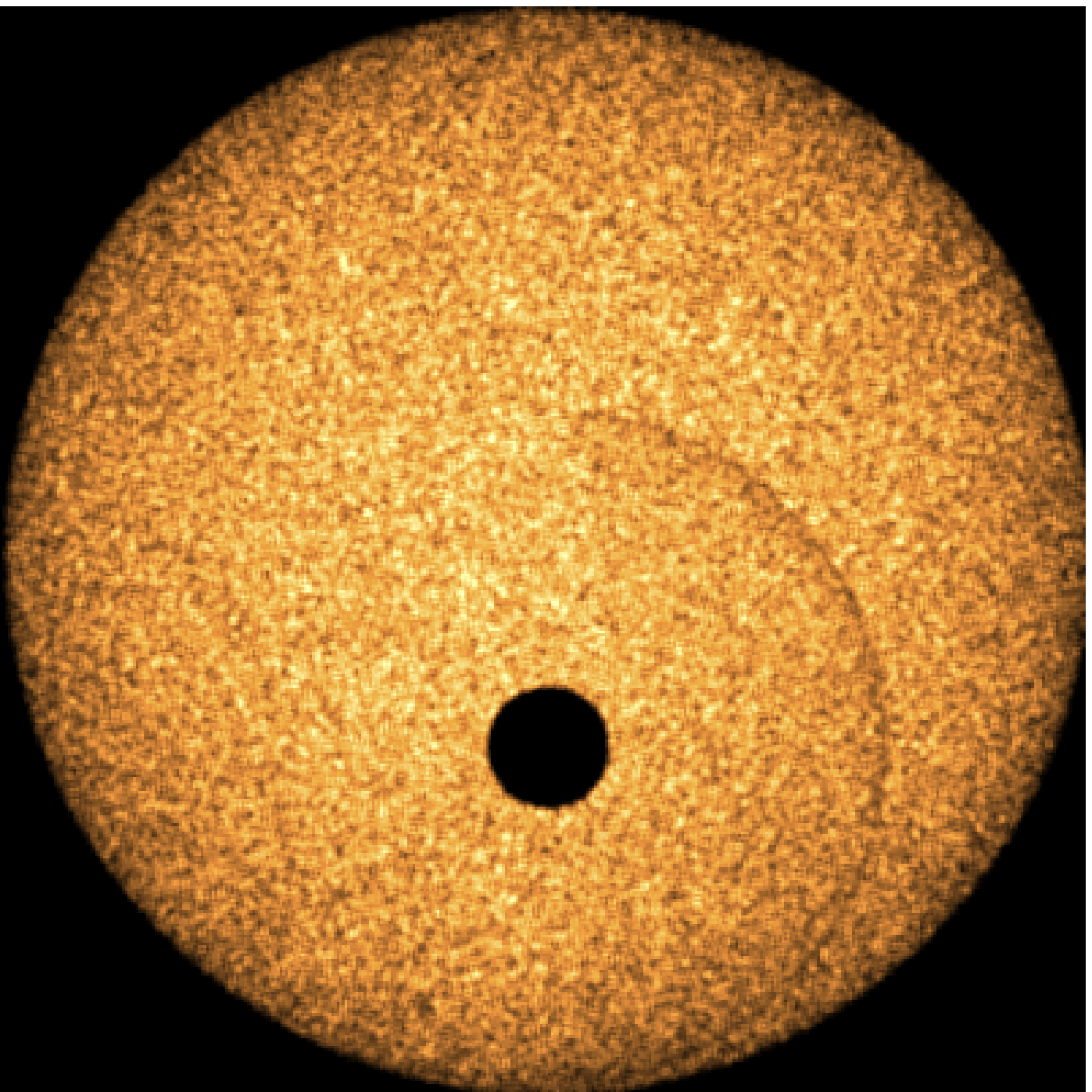} 
      \includegraphics[width=1.5in]{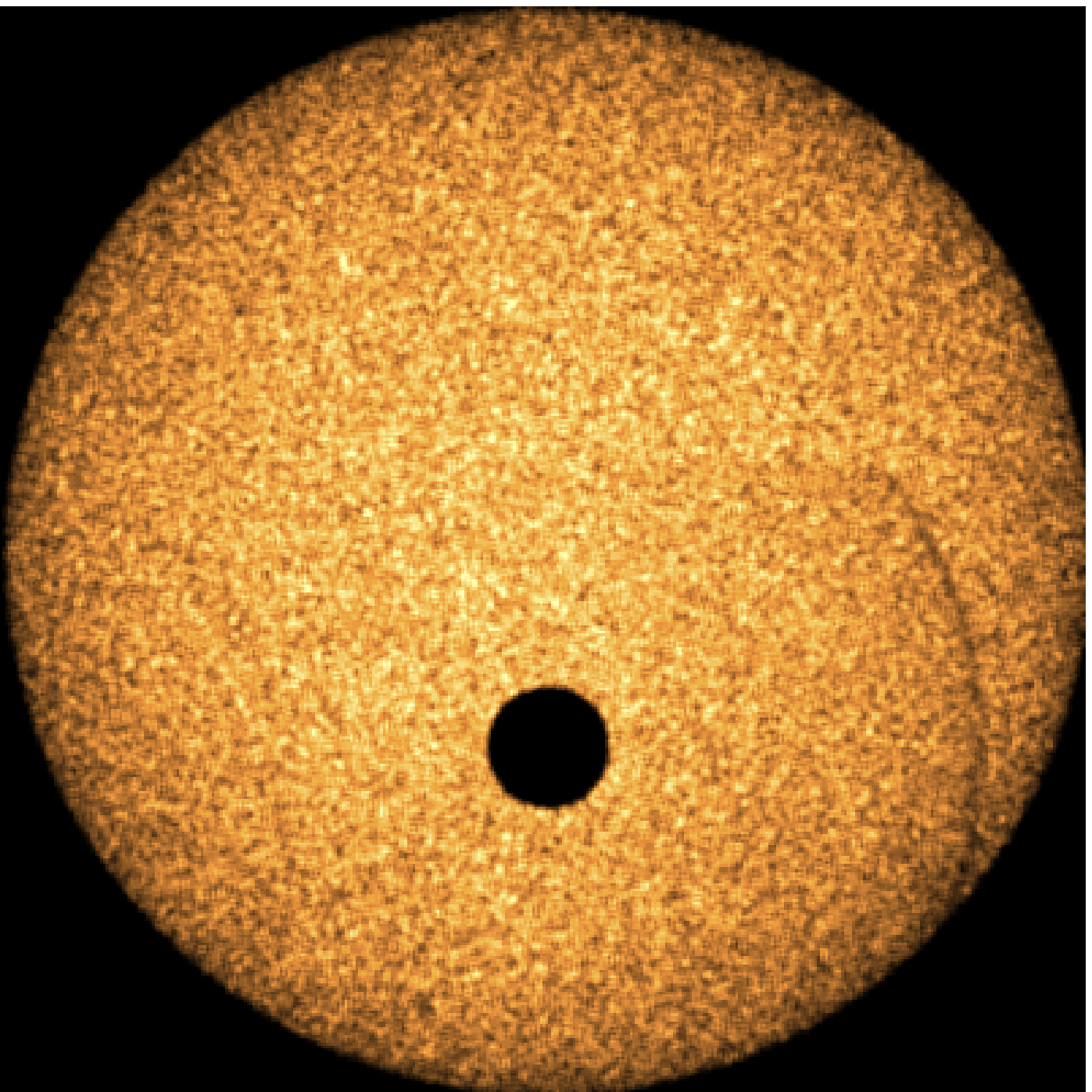}\\
   \caption{Lightcurves and mid-transit images for our simulations. The top panel shows the modelled transits (optical - green; UV - black) with the \textit{HST} observations in red. The black lines represent the results of our simulations. The bottom panel is the mid transit image for each of our models 1A-2B (from left to right respectively). }
   \label{fig:results}
\end{figure*}

\subsection{Monte Carlo Radiation Transfer}
Our simulated transit light curves are produced using a 3D Monte Carlo radiation transfer code  \citep{Wood:1999p822}. The circumplanetary density structure is prescribed on a 3D spherical polar grid (coordinates $r$, $\theta$, $\varphi$) and is externally irradiated with Monte Carlo photon packets with a distribution that reproduces the spatial intensity distribution of a limb-darkened star.  We assume a spherical planet and a limb-darkening law such that the intensity $I$ is given by 
\begin{equation}
	\frac{I(\mu)}{I(0)} = 1-\sum_{n=1}^4a_n(1-\mu^{n/2}), 
\end{equation}
where $\mu = \cos\theta= (1-r^2)^{1/2}, 0\leq r\leq 1$ is the radial distance into the stellar disk normalized to the stellar radius and $I(0)$ is the emergent intensity at the centre of the star \citep{Mandel:2002p814}. The coefficients  $a_n$ are chosen from \cite{Claret:2004p807} to match the $u$-band limb-darkening of the host star. We assume that the material absorbs or scatters radiation out of the line-of-sight with no scattering into the line-of-sight, which is valid for the optical depths required to produce the early-ingress transits. For this letter we assume the bow shock is of uniform density and that the material is static, however, our models are very general and can incorporate any density structure: analytic, tabulated, or from dynamical simulations.
 
\subsection{Analysis of the \textit{HST} observations}
Our goal is to determine the range of shock geometries that can provide both an early-ingress and sufficient optical depth in MgII to account for the absorption seen in the \textit{HST} data \citep{Fossati:2010p838}. The 
\textit{HST} observations (shown in the top panel of Figure \ref{fig:results})  enable us to constrain the allowed shock geometries. The first data point at $\textrm{phase}=0.86$ lies on the continuum of the lightcurve meaning the early ingress is yet to begin. This allows us to find a maximal value for $X_M$. If this value is too large, the dip in emission in our simulated lightcurves will begin too early. We find that we require $X_ M<14R_p$ to ensure a fit to this data point. 

Similarly, the second data point at $\textrm{phase} = 0.92$ provides us with a minimum value for $X_M$. Since this data point no longer lies on the continuum, the bow shock must have started transiting the star. The minimal value for the projected shock distance to fit this data point is $X_M>4.2R_p$ \citep{Lai:2010p808}. We note that the scenario discussed by \cite{Lai:2010p808} where material is accreting from the planet onto the star would result in an accretion stream transiting before the planet and could therefore also produce an early egress.  

For a given density, the depth of the transit lightcurve is determined by the area of shocked material as projected onto the plane of the sky. The data point at $\textrm{phase} = 0.98$ therefore allows us to constrain $\Delta r_M$ and $\Delta\varphi$. If the projected area is too large then the lightcurve will be too deep and if it is too small then the lightcurve will be too shallow. 

If there is shocked material still transiting  after the planet has moved off the stellar disk then we will see a late egress in the transit lightcurve. The data point at $\textrm{phase} = 1.05$  coincides with the optical transit suggesting this is not the case. We therefore require $\varphi_0+\Delta\varphi<90^\circ$  to ensure the near UV transit ends at the same time as the optical transit. 
 
\section{Results}	
We find that an acceptable fit can be achieved for many different plasma temperatures.  These temperatures determine the sound speed and the wind speed. They therefore fix the values of $\varphi_0$ and also the density of the shocked material \citep{Vidotto:2010p809}. 
 
For certain plasma temperatures, however, we are unable to provide a fit to all the data points simultaneously with reasonable shock parameters. For $\textrm{T}=1\times10^6\textrm{K}$, the calculated density is too low and the lightcurve is too shallow. Similarly, for $\textrm{T}=3.93\times10^6\textrm{K}$ the density is too high and the lightcurve is too deep. We therefore choose to concentrate on two temperatures between these values, $\textrm{T}=2\times10^6\textrm{K}$ and $\textrm{T}=2.5\times10^6\textrm{K}$. Because the magnetic field geometry is very unconstrained, for each of these temperatures we present a solution where the planet is embedded in the corona and hence the shock is an ``ahead shock" with $\varphi_0=0^\circ$ and also one where it is immersed in the stellar wind meaning $\varphi_0$ is dependent on the plasma temperature. As illustrative examples we choose $\Delta\varphi=80^\circ$ for models 1A and 2A and $\Delta\varphi=40^\circ$ for models 1B and 2B as these values will ensure the end of the near UV transit will coincide with the end of the optical lightcurve whilst still providing us with a large projected shock area. For all cases we then vary $\Delta r_M$ to find a fit to the observations.

Table \ref{table:results} shows the parameters adopted for our illustrative cases where a fit to all the data points has been found. Figure \ref{fig:results} shows the simulated lightcurves and mid-transit images for our models. From the simulations it is clear that a range of different shock geometries and orientations are able to provide a fit to the data suggesting there is some degeneracy in the solutions. 

Firstly we find that $X_M$ is a degenerate quantity. The value of $X_M$ is determined by the values of $r_M, \varphi_0$ and $\Delta\varphi$ and therefore different combinations of these values can produce the same lateral projected shock extent projected on the plane of the sky. We have found that $X_M=5.5R_p$ provides an acceptable fit to the data.

As the temperature of the stellar plasma increases, the wind and static coronal models predict a larger density at the planetary orbital radius. Therefore the projected area of the shock must decrease to ensure the transit is not too deep and can still fit the data. This area is determined by the angular extent $\Delta\varphi$ and the radial extent $\Delta r_M$. For the cases where $\Delta\varphi=80^\circ$ (models 1A and 2A), a smaller value of $\Delta r_M$ is required to fit the data compared to the cases where $\Delta\varphi=40^\circ$ (models 1B and 2B). This is a consequence of the line-of-sight distance through the shock increasing as $\varphi_0$ decreases, allowing more starlight to be absorbed at the shock front. 
 
In the optical transit the system is completely symmetrical about the mid-transit point, however the addition of a bow shock breaks this symmetry in the near UV transit. This can be seen in the simulated lightcurves as they are not centred around $\textrm{phase}=1$ but offset by an amount proportional to $X_M$. Again if better time sampled observations could be taken, this offset could be used to provide further insight into the stand-off distance between the shock and the planet. 
\section{Discussion}
Our simulations have shown that it is possible to reproduce the \textit{HST} observations of \citet{Fossati:2010p838} by assuming the presence of a bow shock around the planetary magnetosphere. Using models for the stellar corona and wind (with a solar base density) we have calculated the density of MgII in the shocked material. 
 
 \cite{Lai:2010p808} calculated the column density of MgII around WASP-12b to be $\gtrsim 1.4\times10^{13}\textrm{cm}^{-2}$. For this calculation they assume an optical depth $\tau=1$ in the absorption line of MgII, a velocity $v\approx 100\textrm{kms}^{-1}$ and a characteristic length scale $S=3R_p$. From this \cite{Vidotto:2010p809} found the required number density of MgII to be $n_{\rm MgII}\gtrsim400\textrm{cm}^{-3}$ to reproduce the observed early ingress. Using these assumed values we have calculated the extinction cross-section of MgII to be:
\begin{equation}
	\sigma_{\rm MgII}=\frac{\tau}{n_{\rm MgII}S} = 6.5\times10^{-14}\textrm{cm}^2.
	\label{eqn:tau}
\end{equation}

The maximum optical depth $\tau_{\rm max}$ can be then be calculated using Equation \ref{eqn:tau} by setting $S$ to be the maximum path length in the line-of-sight through the shocked material along with the corresponding number density of magnesium, $n_{\rm MgII}$ (from Table \ref{table:results}) for each of our models.  These values are shown in the final column of Table \ref{table:results}.

We have found that it is possible to reproduce the observations with both lower and higher MgII densities and that the resultant shocked material does not need to be optically thick. 

If similar bow shock structures could be observed in other exoplanetary systems, transit observations could be useful to probe the presence of planetary magnetic fields. \cite{Vidotto:2011p803} proposed a list of candidates that could provide signatures of an early-ingress, based on the list of available transiting systems in September 2010. Should UV observations be obtained for these candidates we could apply the model developed here to constrain the allowed geometries and orientations for bow shocks.

\section*{Acknowledgements}
J. Llama acknowledges the support of an STFC studentship. Ch. Helling highlights financial support of the European Community under the FP7 by an
ERC starting grant.
 
\bibliography{bowshocks} 
\bsp
\end{document}